\documentclass[a4paper,fleqn,usenatbib,useAMS]{mnras}

\usepackage{newtxtext,newtxmath}
\usepackage[T1]{fontenc}
\usepackage{ae,aecompl}

\usepackage{graphicx}	% Including figure files
\usepackage{amssymb}	% Extra maths symbols
\usepackage{multicol}        % Multi-column entries in tables
\usepackage{bm}		% Bold maths symbols, including upright Greek
\usepackage{enumitem}

\title[Chaos in the wind of WR~40]{The chaotic wind of WR~40 as probed by \emph{BRITE}\thanks{Based on data collected by the \emph{BRITE-Constellation} satellite mission, designed, built, launched, operated and supported by the Austrian Research Promotion Agency (FFG), the University of Vienna, the Technical University of Graz, the University of Innsbruck, the Canadian Space Agency (CSA), the University of Toronto Institute for Aerospace Studies (UTIAS), the Foundation for Polish Science \& Technology (FNiTP MNiSW), and National Science Centre (NCN).}}

\author[Ramiaramanantsoa et al.]{Tahina Ramiaramanantsoa,$^{1,3}$\thanks{E-mail: tahina@asu.edu} Richard Ignace,$^{2}$ Anthony F. J. Moffat,$^{3}$ \newauthor Nicole St-Louis,$^{3}$ Evgenya L. Shkolnik,$^{1}$ Adam Popowicz,$^{4}$ Rainer Kuschnig,$^{5}$
\newauthor Andrzej Pigulski,$^{6}$ Gregg A. Wade,$^{7}$ Gerald Handler,$^{8}$ Herbert Pablo$^{9}$
\newauthor and Konstanze Zwintz$^{10}$\\\\
$^{1}$ School of Earth and Space Exploration, Arizona State University, 781 E. Terrace Mall, Tempe, AZ, USA 85287-6004\\
$^{2}$ East Tennessee State University, Department of Physics and Astronomy, Johnson City, TN 37614, USA\\
$^{3}$ D\'epartement de physique, Universit\'e de Montr\'eal, CP 6128, Succursale Centre-Ville, Montr\'eal, Qu\'ebec, H3C 3J7\\
$^{4}$ Instytut Automatyki, Politechnika \'Sl\c{a}ska, Akademicka 16, 44-100 Gliwice, Poland\\
$^{5}$ Institute of Communication Networks and Satellite Communications, Graz University of Technology, Infeldgasse 12, 8010 Graz, Austria\\
$^{6}$ Instytut Astronomiczny, Uniwersytet Wroc{\l}awski, Kopernika 11, 51-622 Wroc{\l}aw, Poland\\
$^{7}$ Department of Physics and Space Science, Royal Military College of Canada, Kingston, Ontario K7K 7B4, Canada\\
$^{8}$ Centrum Astronomiczne im. M. Kopernika, Polska Akademia Nauk, Bartycka 18, PL-00-716 Warszawa, Poland\\
$^{9}$ American Association of Variable Star Observers, 49 Bay State Road, Cambridge, MA 02138, USA\\
$^{10}$ Institut f\"{u}r Astro- und Teilchenphysik, Universit\"{a}t Innsbruck, Technikerstrasse 25, A-6020 Innsbruck, Austria\\
}

\date{Accepted  2019 October 10. Received 2019 October 08; in original form 2019 September 04}

\pubyear{2019}

\begin{document}
\label{firstpage}
\pagerange{\pageref{firstpage}--\pageref{lastpage}}
\maketitle

\begin{abstract}
Among Wolf-Rayet stars, those of subtype WN8 are the intrinsically most variable. We have explored the long-term photometric variability of the brightest known WN8 star, WR~40, through four contiguous months of time-resolved, single-passband optical photometry with the \emph{BRIght Target Explorer (BRITE)} nanosatellite mission. The Fourier transform of the observed light-curve reveals that the strong light variability exhibited by WR~40 is dominated by many randomly-triggered, transient, low-frequency signals. We establish a model in which the whole wind consists of stochastic clumps following an outflow visibility promptly rising to peak brightness upon clump emergence from the optically thick pseudo-photosphere in the wind, followed by a gradual decay according to the right-half of a Gaussian. Free electrons in each clump scatter continuum light from the star. We explore a scenario where the clump size follows a power-law distribution, and another one with an ensemble of clumps of constant size. Both scenarios yield simulated light curves morphologically resembling the observed light curve remarkably well, indicating that one cannot uniquely constrain the details of clump size distribution with only a photometric light curve. Nevertheless, independent evidence favours a negative-index power law, as seen in many other astrophysical turbulent media.
 
\end{abstract}


%%%%%%%%%%%%%%%%%%%%%%%%%%%%%%%%%%%%%%%%%%%%%%%%%%

%%%%%%%%%%%%%%%%% BODY OF PAPER %%%%%%%%%%%%%%%%%%

%%%%%%%%%%%%%%%%%%%%%%%%%%%%%%%%%%%%%%%%%%%%%%%%%%%%%%%%%
\section{Introduction}
\label{sec:WR40_Intro}

Hot luminous stars are notorious for their variability, either photometric or spectroscopic, and intrinsic or extrinsic. The most common intrinsic causes include pulsations, rotational modulation, and wind clumps, while extrinsic causes are usually binary-related. Among these stars, the subset of Wolf-Rayet (WR) stars deserves special attention, as they possess the densest stable winds known among luminous hot stars. However, these same dense winds prevent us from exploring their hydrostatic surfaces, where the true driver of the wind lies.  Among WR stars, the intrinsically most variable are the WN8 stars \citep[][and references therein]{1995AJ....109..817A,1998A&A...331.1022M}, with typical variability occurring on timescales of the order of days and with amplitudes at the $\sim$10\% level. However, there is little consensus as to the source of the observed variability, not to mention the lack of clear periodicities which would more easily betray their true source of variability. Nevertheless, \citet{2014MNRAS.440....2M} noted that the cooler subtypes of each of the WN and WC/O sequences tend to be more variable, possibly as a result of their lower surface temperatures. In the scenario of subsurface convection proposed by \citet{2009A&A...499..279C}, the convective layers due to the iron opacity bump in massive stars with cooler surface temperatures lie at somewhat deeper and denser atmospheric levels, where they can produce more driving of surface variability. 

While most previous attempts to define the photometric variability of WR stars in general -- and WN8 in particular -- have been carried out from the ground, we decided to use a space-based observatory to obtain time-dependent precision photometry. Not only does this allow a higher precision than from the ground, but it allows for longer, less gapped observing runs at higher cadence -- all factors that enhance the quality of the photometry and what can be derived from it.  A few attempts have been carried out on several WR stars with the \emph{Microvariability and Oscillations of STars (MOST)} satellite \citep{2005ApJ...634L.109L,2011ApJ...735...34C}, with typical runs lasting about a month at a time, which is barely enough to properly characterize the variability, given the encountered timescales. Therefore, we turned to the \emph{BRIght Target Explorer (BRITE)} constellation of nanosatellites, which would allow contiguous photometric runs from space during up to half a year.

In this investigation we concentrate on only one WN8 star, the WN8h\footnote{``h'' means an observable level of hydrogen in the wind \citep{1996MNRAS.281..163S}, with little effect on the variabilty for WN8 stars \citep{1998A&A...331.1022M}.} star WR~40 (HD 96548; Table~\ref{tab:WR40_StellarParams}).  At $V = 7.7$, it is the brightest WN8 star in the sky and well known for its high level of photometric variability \citep{1980A&A....91..147M,1985A&A...146..307S,1987AJ.....94.1008L,1989MNRAS.240..103B,1994A&A...283..493M,1995AJ....109..817A,1998A&A...331.1022M}. However, the true nature and physical origin of the variability remains highly uncertain, mostly due to the sparsity and short time span of the observations. While the \emph{BRITE} satellites were intended for use on the brightest stars in the sky down to $V = 4$, with some cases reaching $V = 6$, we decided to experiment with an even fainter star, especially one that is known to be strongly variable and isolated on the sky \citep[probably because, like all WN8 stars, it is a runaway;][]{1998A&A...331..949M}. Furthermore, we obtained longer individual exposures in a further attempt to compensate for its faintness. At this magnitude, WR~40 is by far the faintest star that has been successfully targeted so far by \emph{BRITE}.

\begin{table}
\caption{Stellar and wind parameters for WR~40 \citep[from][]{2019A&A...625A..57H}.}
{\normalsize
\begin{center}
\begin{tabular}{l c c c}
\hline
\hline
 Parameter & & & Value  \\
\hline
Spectral type	&	& & WN8h	\\
$\log(L/L_{\sun})$	&	& & $5.91\pm0.15$	\\
$T_{\ast}$~[kK] 	&	& & $44.7\pm3.0$	\\
$R_{\ast}$~[$R_{\sun}$]	&	& & $14.5\pm1.5$	\\
$\dot{M}$~[M$_{\sun}$Myr$^{-1}$]	& &  &	 $63\pm19$	\\
$M_\ast$~[M$_{\sun}$] 	& & & 	 $28$	\\
$\varv_{\infty}$~[$\mathrm{km~s}^{-1}$]	& & &	 $650\pm130$	\\ \hline
\end{tabular}
\end{center}}
\label{tab:WR40_StellarParams}
\end{table}

%%%%%%%%%%%%%%%%%%%%%%%%%%%%%%%%%%%%%%%%%%%%%%%%%%%%%%%%%
\section{\emph{BRITE} photometry of WR~40}
\label{sec:WR40_Obs_BRITE}

\emph{BRITE-Constellation} \citep{2014PASP..126..573W,2016PASP..128l5001P} is a fleet of five nanosatellites operating in low-Earth orbits ($600-790$~km; orbital periods $\sim$$100$~min): \emph{BRITE-Austria (BAb)}, \emph{UniBRITE (UBr)}, \emph{BRITE-Lem (BLb)}, \emph{BRITE-Heweliusz (BHr)}, and \emph{BRITE-Toronto (BTr)}. The mission is geared to exploring the photometric variability of stars brighter than $V\simeq4-6$ in two non-overlapping optical passbands. As such, each $20\times20\times20$~cm nanosatellite hosts a science payload essentially composed of a $3$-cm diameter $f/2.3$ lens-based telescope, an optical filter either operating in the blue regime ($390-460$~nm) or in the red regime ($545-695$~nm), and a $4008\times2672$-pixel KAI-11002M CCD. This configuration offers a relatively large effective unvignetted field of view of $24\degr\times20\degr$ that allows for simultaneous monitoring of up to $\sim$$50$ stars.

\begin{figure*}
\includegraphics[width=18cm]{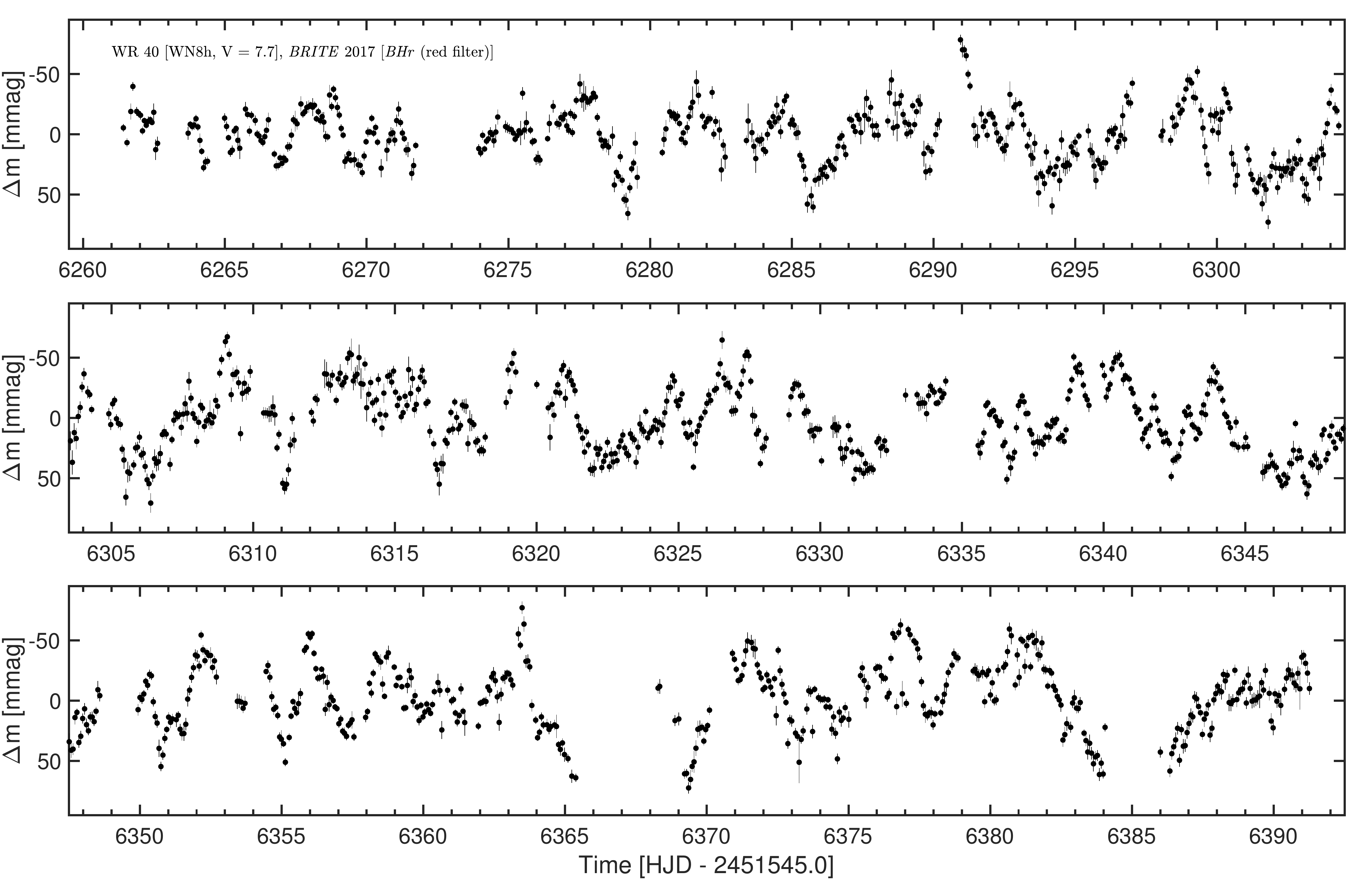}
 \caption{Light curve of WR~40 acquired from four contiguous months of \emph{BHr} photometry in 2017 and binned over each $\sim$$97.1$~min satellite orbit.}
  \label{fig:WR40_BRITE_lc}
\end{figure*}

WR~40 was the coolest of three Galactic H-rich WN stars (the others were WR~22 and WR~24, both with V $\sim$ 6.5) monitored by \emph{BRITE} alongside $47$ other targets during an observing run on the Carina field in 2017. The photometric observations of WR~40 were performed in the red filter by \emph{BHr} between February 21, 2017 and July 01, 2017 (HJD $2457806.408 - 2457936.263$) and consisted of $5$~s exposure snapshots acquired at a median cadence of $20.3$~s during $\sim$$3-29\%$ of each $\sim$$97.1$-min satellite orbit. The overall characteristics of the observations are captured in Table~\ref{tab:WR40_Obs_BRITE_Log}. The time series of raw flux measurements of WR~40 was extracted with the \emph{BRITE} data reduction pipeline described by \citet{2017A&A...605A..26P}, and a multi-dimensional correction of trends exhibited by the flux measurements owing to effects of instrumental origin was applied using the procedure described in \citet{2018MNRAS.480..972R}, who also showed that the post-processing corrections do not affect the intrinsic stellar variability. As the post-decorrelation light curve does not exhibit any obvious signs of short-timescale stellar intrinsic variability within each satellite orbit, we conduct our analyses on the time series of satellite-orbital mean flux measurements depicted in Fig.~\ref{fig:WR40_BRITE_lc}. Despite the faintness of WR~40, clear light variations reaching peak-to-valley amplitudes of $\gtrsim$$0.1$~mag are revealed by the \emph{BRITE} observations. This is not surprising as such large-amplitude light variability has already been reported for WR~40 (see Section~\ref{sec:WR40_Intro}). The strength of these new observations lies in the long time base and moderately high cadence, primarily enabling us to better investigate whether the star truly exhibits coherent (multi-)periodicity or if not, then to reveal the true source of the variability.

\begin{table}
\caption{Summary of the \emph{BRITE} observations of WR~40. The last four entries correspond to values at post-decorrelation stage. $\sigma_{\rm rms}$ is the root mean square mean standard deviation per orbit.}
{\normalsize
\begin{center}
\begin{tabular}{l c c}
\hline
\hline
 Parameter  & & Value  \\
\hline
Satellite									&	& \emph{BRITE-Heweliusz} 	\\
Start -- End dates	 [HJD-2451545.0]			&	& $6261.408-6391.263$	\\
Observing mode$^{\star}$						&	& Chopping				\\
Exposure time [s]							&	& $5$				\\[4pt] 
Total number of data points					&	& $56551$				\\
Number of points per orbit$^{\star\star}$			&	& $37~[8-76]$			\\
Contiguous time per orbit$^{\star\star}$ [min]		&	& $13.2~[2.7-27.8]$		\\
$\sigma_{\rm rms}$ [mmag]					&	& $5.3$				\\ \hline
\end{tabular} 
\end{center}}
{\footnotesize $^{\star}$ The different modes of observations for \emph{BRITE} are described by \citet{2016PASP..128l5001P} and \citet{2017A&A...605A..26P}.\\$^{\star\star}$ Median values. Bracketed values indicate the extrema.}
\label{tab:WR40_Obs_BRITE_Log}
\end{table}

\begin{figure*}
\includegraphics[width=18cm]{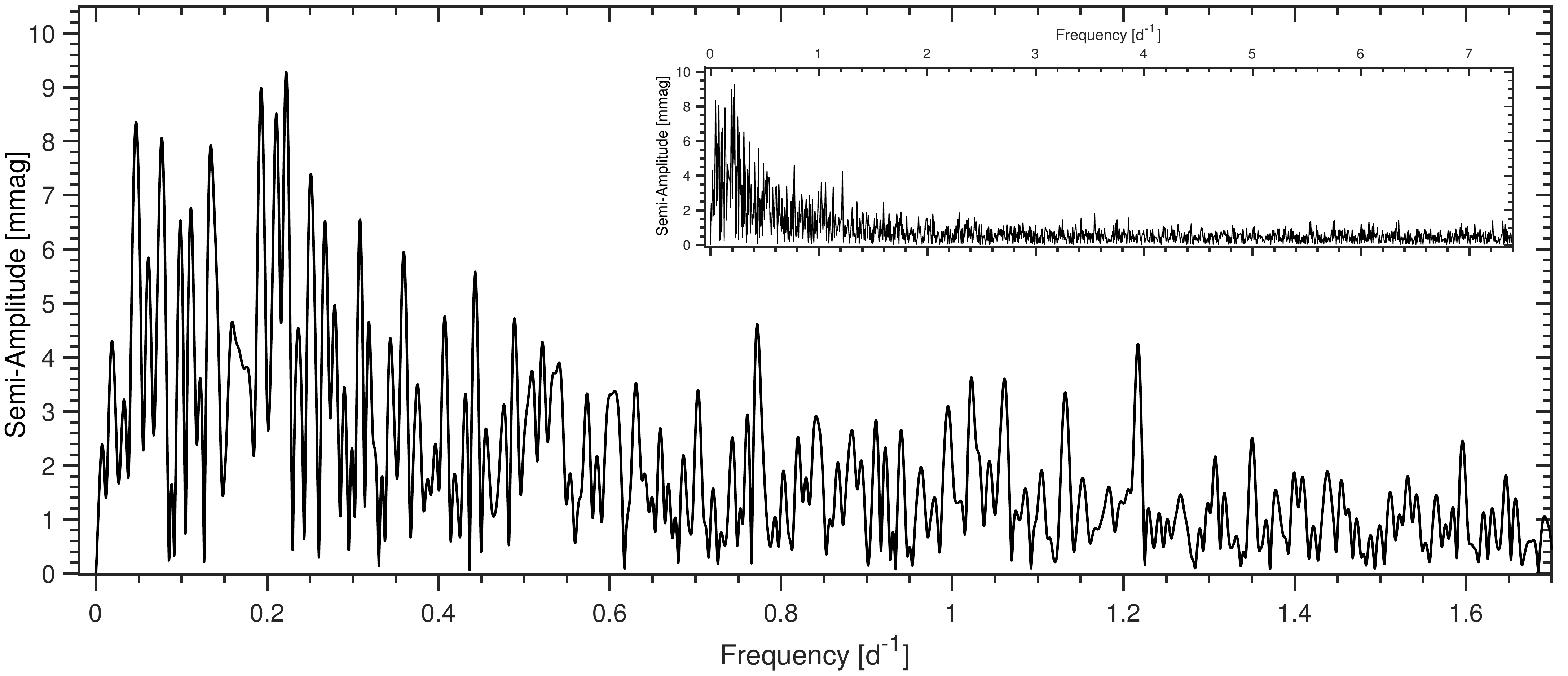}
 \caption{Amplitude spectrum of the \emph{BHr} light curve of WR~40. The inset shows the amplitude spectrum up to the Nyquist frequency at $7.41$~d$^{-1}$.}
  \label{fig:WR40_BRITE_DFT}
\end{figure*}

%%%%%%%%%%%%%%%%%%%%%%%%%%%%%%%%%%%%%%%%%%%%%%%%%%%%%%%%%
\section{Frequency analysis}
\label{sec:WR40_Results_Fourier}

We carried out a period search on the \emph{BRITE} time-resolved photometry of WR~40 using the discrete Fourier transform (DFT) technique implemented in \textsc{Period04} \citep{2005CoAst.146...53L}. As depicted in Fig.~\ref{fig:WR40_BRITE_DFT}, the amplitude spectrum does not exhibit any isolated prominent peak, but reveals an increasing trend towards the low-frequency regime from $\sim$$0.5$~d$^{-1}$ to $\sim$$0.2$~d$^{-1}$, then a decrease below $\sim$$0.2$~d$^{-1}$, resulting in a bump across which there appears to be an ensemble of pronounced peaks.

\begin{figure}
\includegraphics[width=8.4cm]{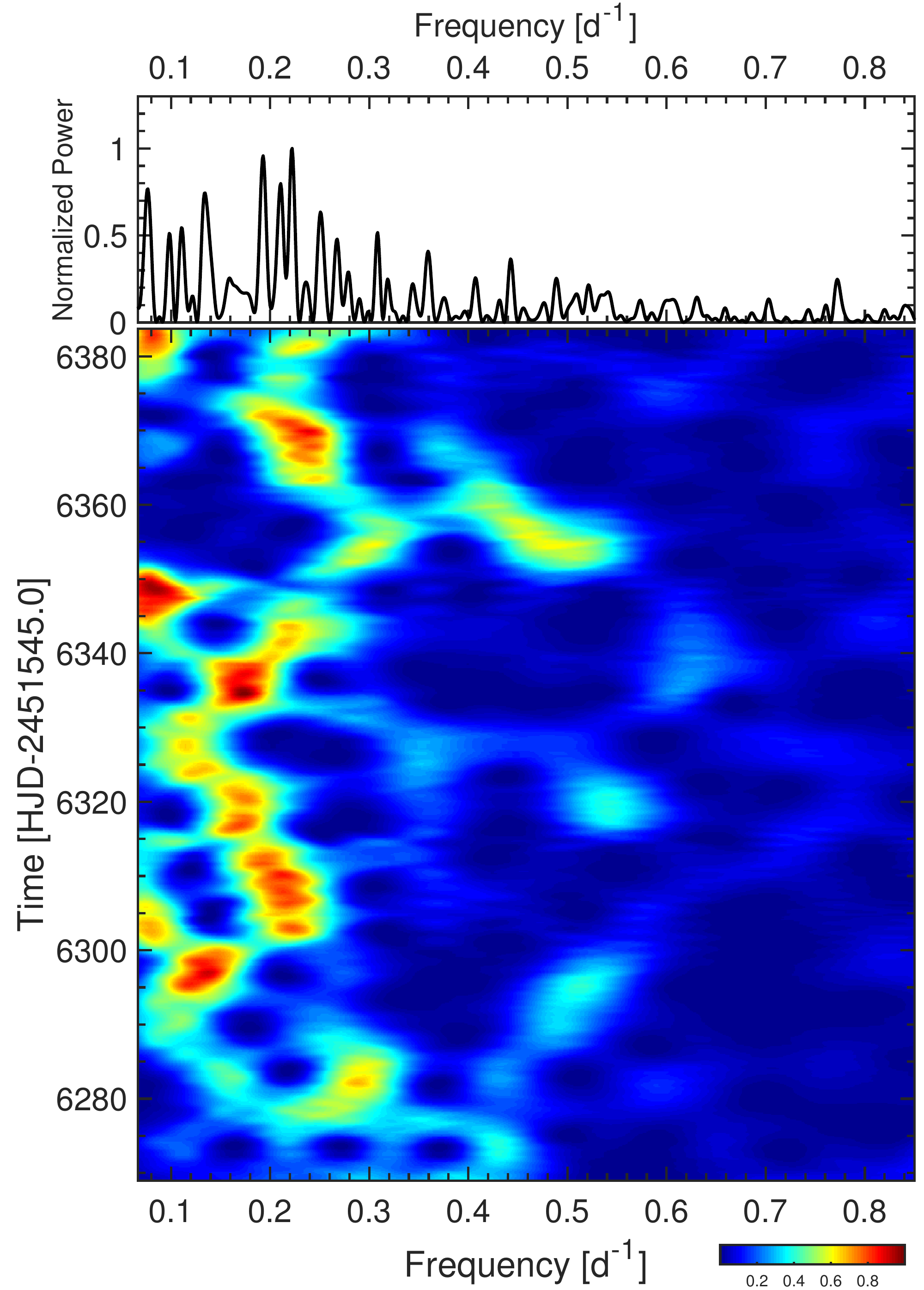}
 \caption{Outcome of a time-frequency analysis of the \emph{BRITE} light curve of WR~40 using a $15$-d sliding, tapered window Fourier transform of the light curve. The $15$-d window size was adopted over other trial window sizes to balance out the constraints in frequency resolution at each step, number of signal cycles covered at each step, and final time base covered by the analysis. The upper panel illustrates the normalized power spectrum of the entire light curve, serving as a guide to locate which signal is excited during which epoch. (Color versions of all figures in this paper are available in the online journal.)}
  \label{fig:WR40_BRITE_TFDiag}
\end{figure}

Given the lack of clearly isolated outstanding signals in the amplitude spectrum of the overall \emph{BRITE} light curve of WR~40, we performed a more detailed time--frequency analysis (Fig.~\ref{fig:WR40_BRITE_TFDiag}) from which it turns out that all the low-frequency signals in the overall DFT are transient, with lifetimes of $\sim$$4-10$~days, and show random triggering as their times of appearance do not follow any organized or coherent behaviour.

%%%%%%%%%%%%%%%%%%%%%%%%%%%%%%%%%%%%%%%%%%%%%%%%%%%%%%%%%
\section{Origin of the variability}
\label{sec:WR40_Results_Interpretation}

The transitory nature of all the dominant low-frequency signals composing the light variations of WR~40 indicates that no obvious manifestation of multiplicity is revealed by our new high-cadence, long time base observations. Particularly, the tallest frequency peak in the observed amplitude spectrum ($\nu = 0.2208\pm0.0116$~d$^{-1}$; $P = 4.53\pm0.24$~d) might be the signal reported by previous investigations that interpreted it to be possibly binary-related \citep{1980A&A....91..147M,1987AJ.....94.1008L}, but now turns out to be part of the forest of randomly triggered signals in the low-frequency regime $\lesssim$$0.5$~d$^{-1}$.

The outcome of the time--frequency analysis is reminiscent of the behaviour of the stochastic component of the light variation in O-type stars as revealed by high-cadence long-term high-precision photometric monitoring from space with \emph{CoRoT (COnvection ROtation and planetary Transits)}, \emph{Kepler}, and \emph{BRITE} \citep{2011A&A...533A...4B,2015ApJ...806L..33A,2017A&A...602A..32A,2018MNRAS.473.5532R,2018MNRAS.480..972R}. That stochastic variability is currently best interpreted as manifestations of randomly-triggered oscillations with finite lifetimes generated in the iron subsurface convection zone and/or internal gravity waves generated at the interface between any convection and radiative zones. However, in the case of those observations of O-type stars, most of the observed light variations arise from the photosphere, whereas here in the case of WR~40 the \emph{BRITE} observations probe variability in the stellar wind. The stochastic nature of the detected variations in the overall light curve of WR~40 then suggests random events of finite lifetime occurring in its wind. At this point, three non-mutually exclusive scenarios can be envisioned.

As a first possibility, if the same types of stochastically-triggered oscillations found in O-type stars are also present in their descendant WR stars, they might propagate into the dense winds of the latter and thus be at least in part responsible for the variability that we detect here. However, given that the mere existence of stochastically-triggered pulsations in WR stars currently remains at the level of pure speculation, and additionally very little (if not nothing at all) is known about how pulsations in WR stars could propagate into their winds, the first scenario remains highly speculative at this point.

Another alternative is variability arising from line-driven instability and wind blanketing. That possibility was recently invoked for O-type stars \citep{2018A&A...617A.121K}, in which line-driven wind instability and wind blanketing would induce relatively large and brief variations of mass-loss rate (sometimes reaching $\sim$$3-5$ times higher than the overall mean on timescales of $\sim$$2.5$~h), which in turn translates into detectable light variations. However, such brief drastic mass-loss rate variations remain to be compared with high-cadence time series of O star mass-loss rate measurements, and the behaviour in WR star winds needs to be investigated.

The third alternative is that scattered starlight by wind clumps themselves might be causing the observed light variations. This is the most plausible scenario, as clumping and its stochastic nature are well-known to occur in WR winds \citep[e.g.][]{1988ApJ...334.1038M,1992PhDT........53R,1994ApJ...421..310M,1999ApJ...514..909L}, in addition to the fact that wind clumping and line-driven instability are found to account for the observed soft X-ray emission in hot stars, either arising from bow shocks preceding the clumps, or from shocks induced by clump-clump collisions \citep[e.g.][]{1980ApJ...241..300L,1997A&A...322..878F,2006MNRAS.372..313O}. In order to test this hypothesis that the \emph{BRITE} observations of WR~40 probe clump-induced scattered light in the stellar wind, we adopt a modeling approach and evaluate the continuum scattered-light variability induced by an ensemble of clumps that come and go over the duration of the \emph{BRITE} observations.  

%%%%%%%%%%%%%%%%%%%%%%%%%%%%%%%%%%%%%%%%%%%%%%%%%%%%%%%%%
\section{Clumped-wind-induced light curve modeling}
\label{sec:WR40_LC_Sims}

%%%%%%%%%%%%%%%%%%%%%%%%%%%%%%%%%%%%%%%%%%%%%%%%%%%%%%%%%
\subsection{Model description}
\label{subsec:WR40_LC_Sims_Model_Descr}

Several theoretical investigations have focused on the scattered light and polarimetric variations from structured stellar outflows in order to relate observed light variations to the physical properties of the clumps, such as mass, size, or outflow velocity \citep[e.g.][]{1995A&A...295..725B,2000A&A...357..233L,2000ApJ...540..412R,2004A&A...426..323B,2009RAA.....9..558L}. Related studies have focused on emission line variability \citep[e.g.][]{1998A&A...335.1003H,1999ApJ...514..909L}. Some of these studies assume that all clumps are identical; others allow for a distribution of clump properties. Here we consider both possibilities. 

Free electrons in the outflowing clumps Thomson scatter light coming from the star. Hence, any individual clump contributes to the observable light variations for a limited interval of time.  Instead of tracking clumps through an outflow, we use a simpler model in which clumps have a characteristic lifetime.  Our simple model takes clumps to emerge from the pseudo-photosphere at peak brightness of scattered light and then fade as $e^{-t^2}$ in time.  On the one hand, this allows for rapid exploration of parameter space.  It is motivated by the fact that many outstanding questions about the stochastic structured flows of massive star winds are unclear.  The state-of-the-art remains at 2D simulations \citep[e.g.][]{2003A&A...406L...1D,2005A&A...437..657D,2018A&A...611A..17S}, in which clump cross-sections can be round, but technically are axisymmetric rings as volumes of revolution about the stellar spin axis, whereas structures are certainly 3D in nature. In our simulation, tracking of individual clump structures would involve assumptions about clump size, their acceleration, their internal expansion, the possibility of collisions between clumps in 3D (and the structures that result), to name a few.  We subsume these issues in terms of a characteristic lifetime in which a clump contributes significantly to scattered light that leads to variability in the continuum light curve measured by \emph{BRITE}.

For the scattered light from an individual optically thin clump, we adopt the geometric configuration used by \citet[][their fig.~1]{1987ApJ...317..290C} and the expression from \citet[][their equation~(2)]{1995A&A...295..725B}:
\begin{equation}
f_{\rm s}/f_\ast = \frac{3\sigma_{\rm T}\,n_{\rm e}\,V_{\rm c}\,{{\cal C}(u,\theta)}}{16\pi\,R_\ast^2}\,(1+\cos^2\theta)\,u^2, 
\end{equation}
where $f_{\rm s}$ is the flux of scattered light from the star, $f_\ast$ is the flux of starlight, $\sigma_{\rm T}$ is the Thomson cross-section, $n_{\rm e}$ is the free-electron
density in the clump, $V_{\rm c}$ is the volume of the clump, $R_\ast$ is the radius of the star, and $u=R_\ast/r$ is the normalized inverse radius, with $r$ the radial location of the clump from the stellar surface.  The angle $\theta$ is the angle of the straight trajectory of the clump relative to the observer's line-of-sight to the star, and the factor ${\cal C}$ is given by:
\begin{equation}
{\cal C}(u,\theta) = \frac{8-\mu\,(1+\mu)\,(1-3\cos^2\theta)}{3\,(1+\mu)\,(1+\cos^2\theta)},
\end{equation}  
in which $\mu = \sqrt{1-u^2}$.

Our approach to modeling the characteristic light variations observed in WR 40 involves a few modest assumptions regarding the limit $u\rightarrow1$, clump lifetime, and clump mass distribution as detailed in the following sections.

%%%%%%%%%%%%%%%%%%%%%%%%%%%%%%%%%%%%%%%%%%%%%%%%%%%%%%%%%
\subsubsection{Behaviour near the hydrostatic core surface}
\label{subsubsec:WR40_LC_Sims_u=1}

Approaching the stellar photosphere ($u\rightarrow1$), the fraction of scattered light by a clump reduces to
\begin{equation}
f_{\rm s}/f_\ast = \frac{\sigma_{\rm T}\,n_{\rm e}\,V_{\rm c}}{2\pi\,R_\ast^2},
\end{equation}
%really
which holds for all $\theta$ trajectories of the clumps.  Since the mass of the clump relates to volume via $m_{\rm c}=\mu_{\rm e}\,m_{\rm H}\,n_{\rm e}\,V_{\rm c}$, where $\mu_{\rm e}$ is the mean molecular weight per free electron, the expression for scattered light can be recast in terms of mass as
\begin{equation}
m_{\rm c} = \frac{2\pi\,\mu_{\rm e}\,m_{\rm H}\,R_\ast^2}{\sigma_{\rm T}}\,f_{\rm s}/f_\ast.
\label{eq:clump_mass}  
\end{equation}

\noindent Note that for WR~40, \citet{2019A&A...625A..57H} find $X=23\%$ for the mass fraction of H.  Assuming a composition that is entirely ionized H and singly-ionized He, the mean molecular weight is $\mu_{\rm e}=2.4$ for WR~40.  As a result, provided with a composition and state of ionization, and with an estimate for the stellar radius, the clump mass for thin electron scattering can be related to an observable in the form of the brightness fluctuations of the source.

%%%%%%%%%%%%%%%%%%%%%%%%%%%%%%%%%%%%%%%%%%%%%%%%%%%%%%%%%
\subsubsection{Clump lifetime}
\label{subsubsec:WR40_LC_Sims_ClumpLifetime}

Another key parameter is the lifetime of a clump. Here, lifetime refers to how individual clumps brighten and fade over time.  This naturally occurs as a clump moves outward from the star.  If a clump evolves at constant solid angle, then
$f_{\rm s}/f_\ast \propto u^2$.  Thus when a clump becomes visible at the pseudo-photosphere (taken to be at an optical depth of 1, located at $2.5R_\ast$), it appears at maximum brightness and then fades according to $u(t)$.  With the selection of a velocity law for the clump, one could solve for $u(t)$.  Instead, we adopt the following prescription as an adequate approximation for how a clump fades with time, using
\begin{equation}
f_{\rm s}/f_\ast = f_0/f_\ast\,H(\Delta t)\,e^{-\Delta t^2/\tau^2},
\label{eq:clump_fading}    
\end{equation}
where $f_0$ is the scattered light at the pseudo-photosphere,
$\tau$ is the characteristic lifetime of the clump as it grows dimmer, $\Delta t$ is the travel time of the clump following its appearance at the pseudo-photosphere, and $H(\Delta t)$ is the Heaviside function, which is 0 for $\Delta t<0$ and 1 for $\Delta t\ge 0$.  The parameter $\tau$ is a free parameter of the model that helps determine the smoothness in the light fluctuations,
for a given ensemble of clumps.  Note that the scattered light contributed by an individual clump drops to $1/e$ when $\Delta t = \tau$.  Because the fading of this contribution is the declining half of the Gaussian, the half-width at half maximum is $\Delta t_{1/2}=0.83\tau$.

%%%%%%%%%%%%%%%%%%%%%%%%%%%%%%%%%%%%%%%%%%%%%%%%%%%%%%%%%
\subsubsection{Clump mass distribution}
\label{subsubsec:WR40_LC_Sims_MassDistr}

As previously mentioned, we explore two scenarios: one in which the clump mass follows a power-law distribution (Case A hereafter), and another one where all clumps have the same mass (Case B hereafter).

Model Case A is founded on potential observational evidence in support of a distribution of clump mass, with $N(m_{\rm c}) \propto m_{\rm c}^{-3/2}$ \citep{1994RvMA....7...51M,1994Ap&SS.221..137R}. For our assumption of optically thin scattering, $f_{\rm s}\propto m_{\rm c}$, and so the scattered-light distribution is likewise a power law with a $-3/2$ exponent. This scenario requires as input the fractional flux of scattered light by the most massive clump, $f_{\rm max}/f_\ast$. This sets the amplitude of light variations in the simulated light curves. Then equation~\eqref{eq:clump_mass} provides for how this free parameter relates to the mass of the largest clump, $m_{\rm max}$.

It would seem that another free parameter would be the number of clumps in the simulation. However, this is not actually a free parameter, because the outflow is constrained by the known (or approximate) stellar mass-loss rate $\dot{M}$. Once $f_{\rm max}/f_\ast$ is chosen, $m_{\rm max}$ is determined. Combined with the distribution of clumps and the duration of the observations for which the simulation is seeking to match the light curve characteristics, the number of clumps is constrained by the value of $\dot{M}$.

For a distribution of clump masses varying as $m^{-3/2}$, the average clump mass $\bar{m}_{\rm c}$ is the geometric mean for the product of the maximum and minimum clump masses in the simulation, such that the mass-loss rate is given by:
\begin{equation}
\dot{M} = \bar{m}_{\rm c}\,\dot{{\cal N}} =  \sqrt{m_{\rm max}\,m_{\rm min}}\,
    \frac{N_T}{T},
    \label{eq:mdot}
\end{equation}
where $\dot{{\cal N}}$ is the rate at which clumps become visible in the flow, $T$ is the time interval of the simulation (corresponding to the duration of the observations), and $N_T$ is the total number of clumps over that period of time.
That expression contains two known quantities -- $\dot{M}$ and $T$.  The maximum clump mass is taken as uniquely determined by $f_{\rm max}/f_\ast$.  This leaves the minimum clump mass and total number of clumps as unknowns.  However, these two quantities are not independent. For example, choosing the minimum mass fixes $\bar{m}_{\rm c}$, and then $N_T$ would be given by the relation with mass-loss rate.

Our simulation discretizes clumps according to clump masses as follows.  Over the duration $T$ of the simulation, it is assumed that there is only one clump with the largest mass, $m_1 = m_{\rm max}$.  Then there are two clumps of the next biggest mass, then three, and so on. We denote this count for the clump scale size as ``$j$'', which ranges from $1$ to $J$, with $J$ being the number of clumps having the smallest mass.  Given that the number of clumps is $N \propto m^{-3/2}$, our discretization approach implies that $m_{\rm j} = j^{-2/3}$.  As $j$ increases, the clump mass decreases as the 2/3 power of this parameter.

The discretization scheme also implies that the total number of clumps is $N_T = 0.5\,J\,(J+1)$.  With $J\gg 1$, which will be the case for our simulations, $N_T \approx J^2/2$.  But we also know that $m_J = J^{-2/3}$.
Combining these relations into equation~\eqref{eq:mdot} provides the value of $J$ in terms of other given quantities, with
\begin{equation}
J \approx \left(\frac{2\dot{M}\,T}{m_{\rm max}}\right)^{3/5}.
\end{equation}
\noindent With $J$ determined, both the minimum mass $m_{\rm min} = m_J$ and the total number of clumps $N_T$ are likewise determined.  Consequently, our simulations to reproduce the observed light curve of WR~40 -- in its characteristics, not as a fit -- have only two free parameters:  $f_{\rm max}/f_\ast$ and $\tau$.

For Case B, again the lifetime and the maximum flux of scattered light are the free parameters.  However, there is no distribution of clump masses as all clumps are identical for Case B.  Changing the maximum flux still changes the mass of each clump and thus the number of clumps in the simulation necessary to support the mass-loss rate.

%%%%%%%%%%%%%%%%%%%%%%%%%%%%%%%%%%%%%%%%%%%%%%%%%%%%%%%%%
\subsubsection{Clump birth times}
\label{subsubsec:WR40_LC_Sims_ClumpGen}

Owing to the statistical nature of the simulation, it is not possible to fit the actual light curve, given the underlying assumption that the mass loss is clumped  and involves some level of randomness.  A stochastic element to the modeling enters through the assignment of times when a clump is taken to become visible at the pseudo-photosphere.  A random number generator is used to uniformly sample the time when a clump emerges from the pseudo-photosphere.  At the moment of emergence, the clump is at its brightest in scattered starlight.  At any given time, there is a typical number of clumps, $\bar{N}(t)$ that are relatively bright such that $\Delta t < \tau$, as given by $\bar{N} = N_T \times (\tau/ T)$.

%%%%%%%%%%%%%%%%%%%%%%%%%%%%%%%%%%%%%%%%%%%%%%%%%%%%%%%%%
\subsection{Model Results}
\label{subsec:WR40_LC_Sims_Results}

\begin{figure*}
\includegraphics[width=16.5cm]{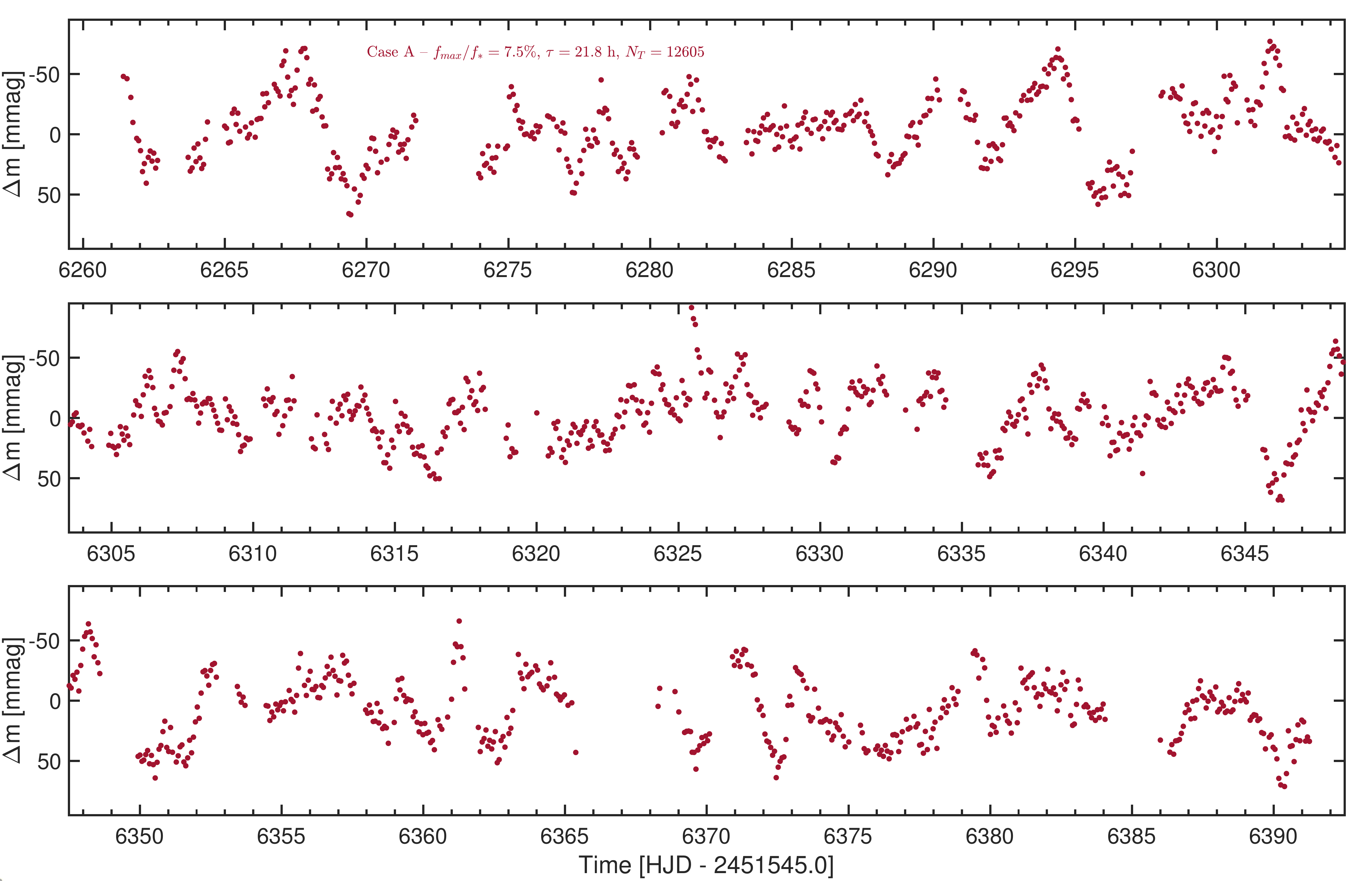}
 \caption{Example of model light curve for Case A.}
  \label{fig:WR40_Simul_CaseA_LC}
\end{figure*}

\begin{figure*}
\includegraphics[width=16.5cm]{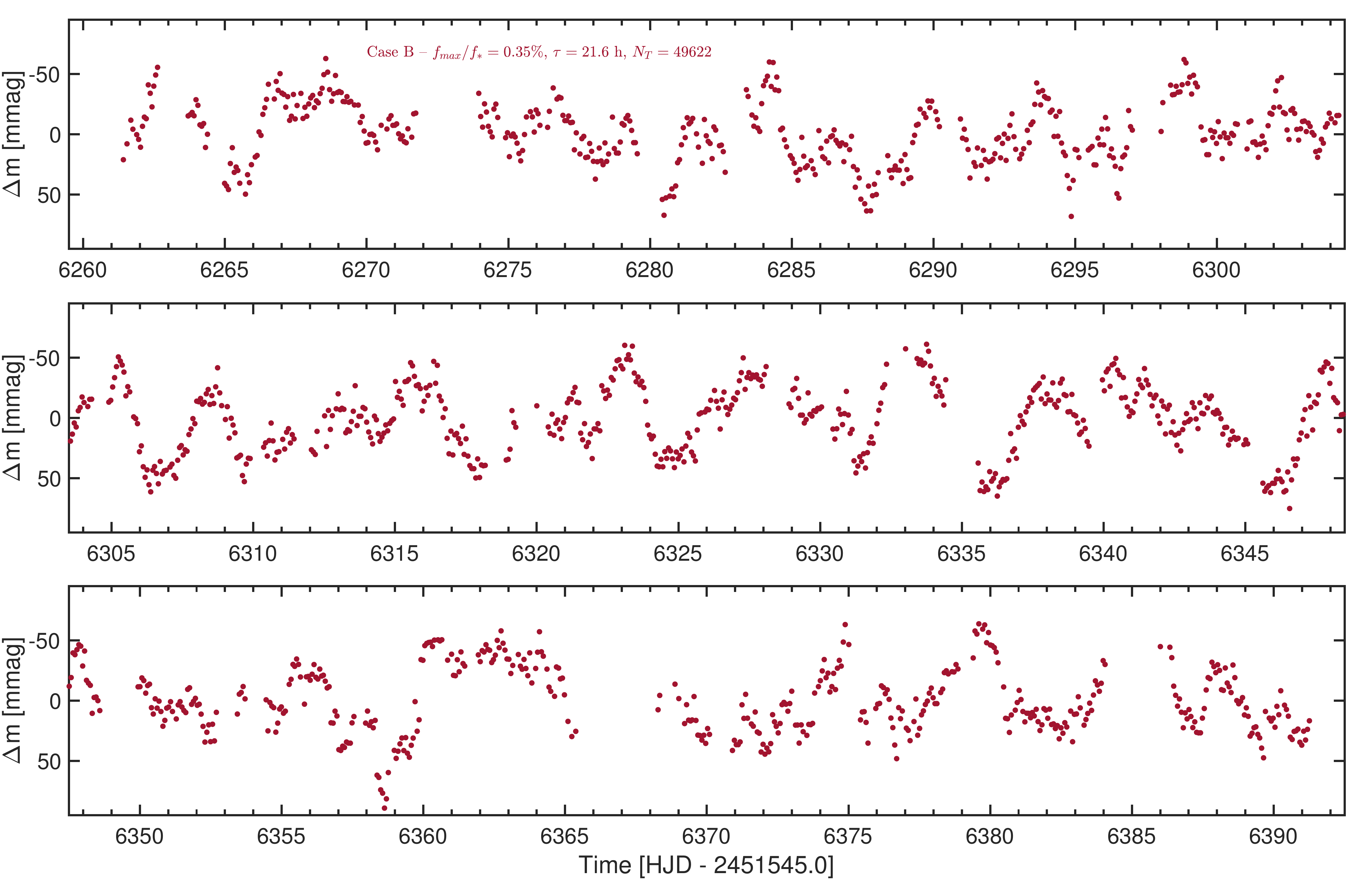}
 \caption{Same as in Fig.~\ref{fig:WR40_Simul_CaseA_LC} but for Case B.}
  \label{fig:WR40_Simul_CaseB_LC}
\end{figure*}

\begin{table}
\caption{Model parameters for the synthetic light curves depicted in Figs~\ref{fig:WR40_Simul_CaseA_LC}~and~\ref{fig:WR40_Simul_CaseB_LC}.}
\centering
{\normalsize
\begin{center}
\begin{tabular}{l c c c c}
\hline
\hline
Parameter	&	& &	Case A	&	Case B\\
\hline
$f_{\rm max}/f_\ast$~[\%]	& & 	&	7.5 & 0.35 	\\
$J$	&	& &158	&	--\\
$N_T$	&	& &	12605&	49622\\
$m_{\rm max}$~[g]	&	& &	$1.8\times10^{25}$&$8.6\times10^{23}$	\\
$m_{\rm min}$~[g]	&	& &	$3.4\times10^{-2}$&	$8.6\times10^{23}$\\
$\tau$~[h]	&	& &	21.8 &	21.6\\
\hline
\end{tabular}
\end{center}}
\label{tab:WR40_Simul_Params}
\end{table}

\begin{figure*}
\includegraphics[width=18cm]{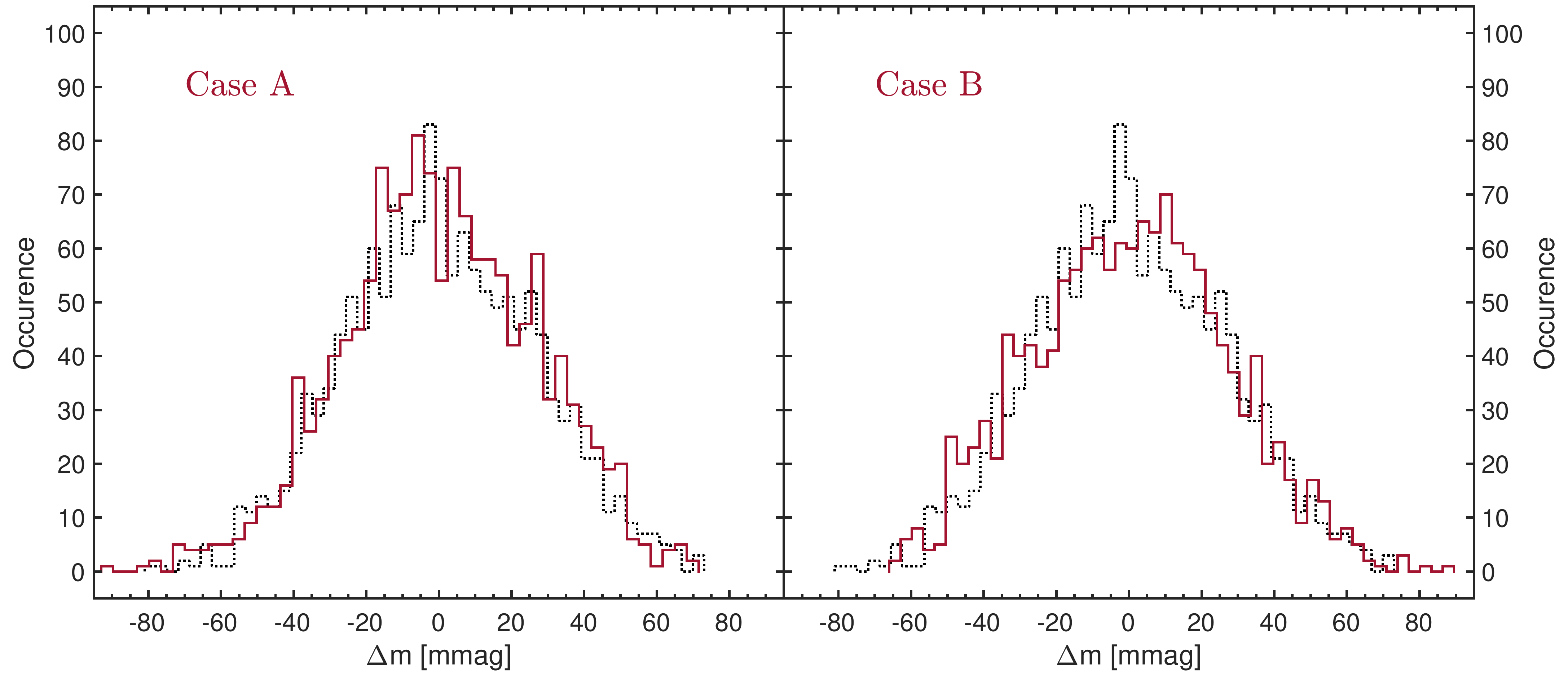}
 \caption{Histograms of the model light curves illustrated in Figs~\ref{fig:WR40_Simul_CaseA_LC}~and~\ref{fig:WR40_Simul_CaseB_LC} (red) compared to that of the \emph{BRITE} observations (black).}
  \label{fig:WR40_Simul_Hist}
\end{figure*}

\begin{figure*}
\includegraphics[width=18cm]{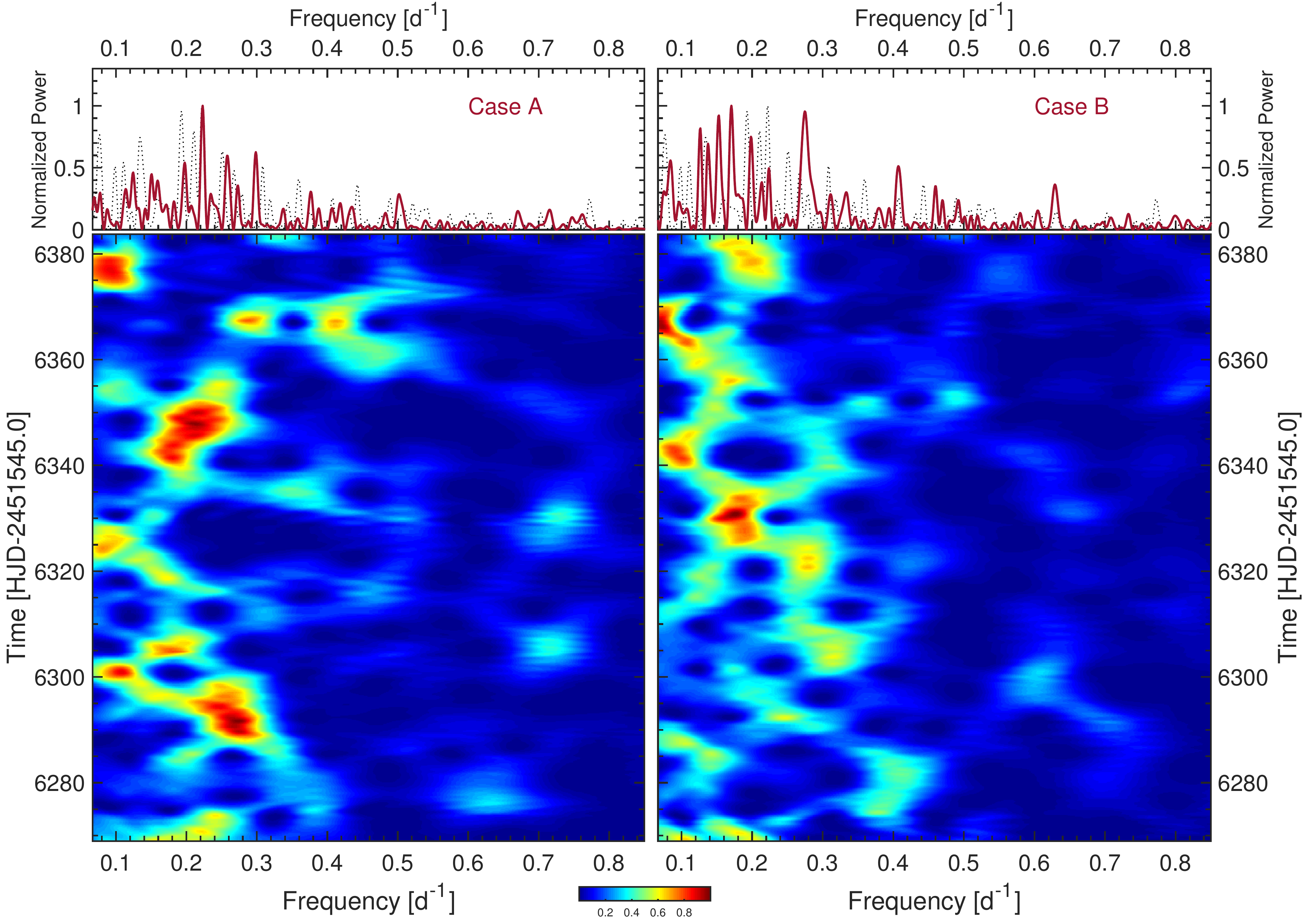}
 \caption{Time-frequency diagrams resulting from a $15$-d sliding window Fourier transform of the model light curves shown in Figs~\ref{fig:WR40_Simul_CaseA_LC}~and~\ref{fig:WR40_Simul_CaseB_LC}. In the top panels, the solid red curves are the periodograms of the simulated light curves whereas the dotted black curve is that of the observations.}
  \label{fig:WR40_Simul_TFDiags}
\end{figure*}

We perform all our simulations adopting the stellar and wind parameters of WR~40 listed in Table~\ref{tab:WR40_StellarParams}. Figs~\ref{fig:WR40_Simul_CaseA_LC}~and~\ref{fig:WR40_Simul_CaseB_LC} show examples of light curve simulations for Case A and Case B, respectively, with the corresponding model parameters listed in Table~\ref{tab:WR40_Simul_Params}. 

As already noted in the previous section, the only free parameters in our modeling approach aiming to reproduce the characteristics of the observed \emph{BRITE} light curve of WR~40 are the maximum fractional flux of scattered light induced by the biggest clump ($f_{\rm max}/f_\ast$), and the characteristic lifetime $\tau$ of a clump.

To identify reasonably good parameters for the model in order to match the characteristics of the observed light curve, we initially used a visual inspection, and for reasonable parameters, we explored more carefully the match through a 
comparison of histograms between the observed and simulated light curve amplitudes. Fig.~\ref{fig:WR40_Simul_Hist} displays a comparison between the histograms of the \emph{BRITE} data and those of the simulations. The match is quite good. When $f_{\rm max}/f_\ast$ is too large, the histogram for the simulated light curve would be too broad; when $f_{\rm max}/f_\ast$ is too small, the histogram is too narrow.  The matching of histograms is not sufficient to identify a uniquely good model.  We find that a range of clump lifetimes yields reasonably similar histograms for light variations.  However, values of $\tau$ that are too short result in indistinct peaks and troughs in the simulation.  When $\tau$ is too long, the peaks and troughs are too well-defined, or too smooth.

To illustrate these effects, Fig.~\ref{fig:WR40_Simul_ParamStudy} shows other examples of Case A models along with their histograms. This parameter study reveals that varying $f_{\rm max}/f_\ast$ has consequences for the value of $m_{\rm max}$, which then impacts the value of $J$, $N_T$, and $m_J$ for the simulation. By contrast $\tau$ is completely independent of any other factors for the simulation.  

Lastly, as a final validation of the modeled light variations we inspected their power spectra and time-frequency diagram behaviour (Fig.~\ref{fig:WR40_Simul_TFDiags}). The range of prominent frequencies detected in the synthetic light curves ($\sim$$\left[0.05; 0.4\right]$~d$^{-1}$), as well as their transiency and their lifetimes as revealed by the time-frequency analyses, are in good agreement with those of the observations. Naturally, these signal lifetimes in the sense of the Fourier analysis, which represent how long a detected sinusoid at a given frequency lasts in the light curve, are different from the clump lifetimes defined in equation~\eqref{eq:clump_fading} which represent the timescale on which the brightness of a single clump drops by a factor $e$.

%%%%%%%%%%%%%%%%%%%%%%%%%%%%%%%%%%%%%%%%%%%%%%%%%%%%%%%%%
\section{Discussion}
\label{sec:WR40_Discussion}

The relatively low total number of clumps yielded by the simulations are noteworthy, as they imply an average of $88$ clumps at any given time for the Case A model and $344$ clumps for the Case B model. As mentioned in Section~\ref{sec:WR40_Results_Interpretation}, soft X-ray emission in hot stars are best interpreted as manifestation of clump-induced shocks. In the case of WR~40, \citet{2005A&A...429..685G} did not detect any significant X-ray emission, and suggested on the basis of a modeling of the opacity of its wind that such non-detection could be due to the fact that the stellar wind optical depth is large enough to block X-rays. Additionally, the low number of clumps suggested by the outcome of our modeled optical light curves could imply rarer clump-induced shocks, hence could additionally account for the non-detection results reported by \citet{2005A&A...429..685G}.

With regard to timescales, both model Case A and Case B appear to favor a clump characteristic lifetime of $\sim$$22$~h, which is much shorter than the dominant variability timescale of $\sim$$4.5$~d revealed by the Fourier analysis of both the observed and the modeled light curves. This could be the consequence of strong nesting effects, which inevitably occur as the contributions of several clumps to the overall light variability yield a broader trend in the light curve that the Fourier analysis interprets as low-frequency variability. On the other hand, the $22$-h clump characteristic lifetime could also be compared with the wind flow time. Taken to be the ratio between the radius of the pseudo-photosphere ($2.5R_\ast$) and the terminal wind speed, the wind flow time is $\sim$$11$~h for WR~40. This implies that the number of clumps per flow time is roughly half the average number of clumps (at any given time) implied by the simulations.

%%%%%%%%%%%%%%%%%%%%%%%%%%%%%%%%%%%%%%%%%%%%%%%%%%%%%%%%%
\section{Conclusion}
\label{sec:WR40_Conclusion}

The $\sim$$4$-month high-cadence \emph{BRITE} photometric observations monitoring the behaviour of the dense wind of the WN8h star WR~40 revealed a high level of light variability mostly composed of stochastically-triggered, low-frequency transient signals. We have successfully simulated the observed light variations using a simple model of discrete clumps that comprise the whole stellar wind and form somewhere below the visible part of the wind \citep[most likely at the hydrostatic core surface, as observed in O stars;][]{2018MNRAS.473.5532R}. However, the fact that we can simulate the light curve with or without a power-law distribution for the clump masses points towards two important conclusions. On one hand, it means that the light curve is not sensitive to the exact details of the clumping, only that clumping of a substantial (probably the whole) part of the wind is necessary. On the other hand, it indicates that the light curve does not constrain what the clump mass distribution actually is. The same degeneracy was seen in the line-profile variations in time-series spectra of several WR stars \citep{1999ApJ...514..909L}, where the variable subpeaks on the lines are examined. However, other analyses of those same subpeaks suggested that a power law might in fact be present for at least all subpeaks within an order of magnitude up to the brightest subpeak \citep{1994RvMA....7...51M,1994Ap&SS.221..137R}. Evidently, each subpeak need not refer to a single clump, but may be a combination of many. Intensive spectroscopic monitoring of WR~40 (e.g. its He~{\sc ii}~$\lambda5411$ emission line) performed in parallel with time-dependent photometric observations such as the ones provided by \emph{BRITE} might help constrain the extent of such nesting effects by direct comparison of the number of subpeaks found in the line profile variations and the amplitudes of the bumps observed in the light variations at different times. In any case, the fact remains that in the context of compressible turbulence, such nesting effects are common and do not deter from examining the power law, which in this case follows that expected for compressible turbulence. Such behaviour of many other types of astrophysical plasmas/gas is known, including e.g. the cloudlets in giant molecular clouds and the intergalactic medium, but also the mere existence of an initial mass function for stars and even galaxies.

Our results here for just one star, albeit one with a strong wind and a high level of variability, may have far-reaching consequences. Since all hot-star winds appear to be clumped, the description that best fits that clumping could very well be a scenario of compressible turbulence,  potentially driven by surface perturbations from subsurface convection in the inner part of the wind and by line-driven instability further out in the wind.

%%%%%%%%%%%%%%%%%%%%%%%%%%%%%%%%%%%%%%%%%%%%%%%%%%%%%%%%%
\section*{Acknowledgements}
We thank the reviewer, Achim Feldmeier, for his insightful suggestions on key aspects of our paper. TR and ELS acknowledge support from the NASA APRA program (NNH16ZDA001N-APRA). RI acknowledges support by the National Science Foundation under Grant No. AST-1747658.  AFJM and NSL are grateful for financial aid from NSERC (Canada). APo was responsible for image processing and automation of photometric routines for the data registered by the \emph{BRITE}-nanosatellite constellation, and was supported by statutory activities grant SUT 02/010/BKM19 t.20. APi acknowledges support from the NCN grant no. 2016/21/B/ST9/01126. GAW acknowledges support from NSERC (Canada) in the form of a Discovery Grant. GH thanks the Polish National Center for Science (NCN) for support through grant 2015/18/A/ST9/00578.

%%%%%%%%%%%%%%%%%%%%%%%%%%%%%%%%%%%%%%%%%%%%%%%%%%



%\bibliographystyle{mnras}


%%%%%%%%%%%%%%%%%%%%%%%%%%%%%%%%%%%%%%%%%%%%%%%%%%

%%%%%%%%%%%%%%%%% APPENDICES %%%%%%%%%%%%%%%%%%%%%

%%%%%%%%%%%%%%%%%%%%%%%%%%%%%%%%%%%%%%%%%%%%%%%%%%%%%%%%%
%%%%%%%%%%%%%%%%%%%%%%%%%%%%%%%%%%%%%%%%%%%%%%%%%%%%%%%%%
\appendix
%%%%%%%%%%%%%%%%%%%%%%%%%%%%%%%%%%%%%%%%%%%%%%%%%%%%%%%%%
%%%%%%%%%%%%%%%%%%%%%%%%%%%%%%%%%%%%%%%%%%%%%%%%%%%%%%%%%

%%%%%%%%%%%%%%%%%%%%%%%%%%%%%%%%%%%%%%%%%%%%%%%%%%%%%%%%%
\section{Light curve modeling -- Parameter study}
\label{sec:WR40_Simul_ParamStudy}

\begin{figure*}
\includegraphics[width=18cm]{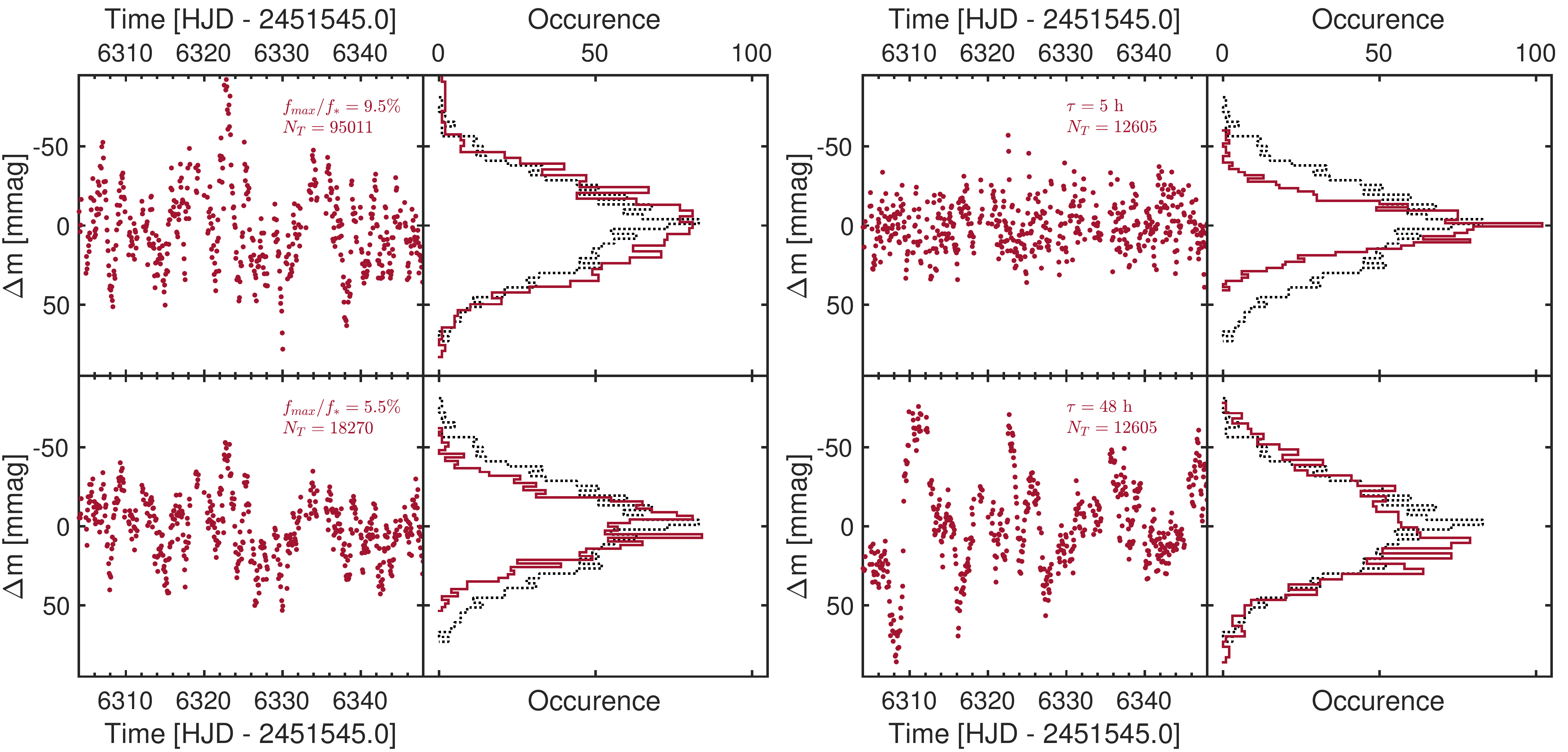}
 \caption{Model parameter study. The first two columns show excerpts of simulated light curves ($44$~d) for two $f_{\rm max}/f_\ast$ values with $\tau$ fixed at $24$~h along with histogram comparisons between the entire WR~40 data (black) against the full simulated light curves (red). The last two columns show the same type of information as in the first two, but for two $\tau$ values with $f_{\rm max}/f_\ast=7.5\%$.}
  \label{fig:WR40_Simul_ParamStudy}
\end{figure*}

%%%%%%%%%%%%%%%%%%%%%%%%%%%%%%%%%%%%%%%%%%%%%%%%%%

% Don't change these lines
\bsp	% typesetting comment
\label{lastpage}
\end{document}